\documentclass[a4paper,twocolumn,12pt]{article}
\usepackage[spanish]{babel}
\usepackage{ae}
\usepackage[latin1]{inputenc} 
\usepackage[T1]{fontenc} 
\usepackage{layout}
\usepackage{array}
\usepackage{graphicx}
\usepackage{amsmath}
\usepackage{multicol}
\usepackage{rotating}  
\hyphenrules{nohyphenation} 

\usepackage[]{cite}   
\usepackage{anysize}  
\usepackage{endnotes} 

\usepackage{tikz,fp,ifthen,fullpage}
\usetikzlibrary{arrows,snakes,backgrounds}
\usepackage{pgfplots}
\usetikzlibrary{shapes,trees}
\usetikzlibrary{calc,through,backgrounds,decorations}
\usetikzlibrary{decorations.shapes,decorations.text,decorations.pathmorphing,backgrounds,fit,calc,through,decorations.fractals}
\usetikzlibrary{fadings,intersections}
\usetikzlibrary{patterns}
\usetikzlibrary{mindmap}
\marginsize{1.5cm}{1.0cm}{1.5cm}{2cm}

\begin{document}

\renewcommand{\tablename}{Tabla}
\renewcommand{\abstractname}{}
\renewcommand{\thefootnote}{\arabic{footnote}}

\title{
        {\LARGE Velocidades: media, promedio e instantánea en el movimiento uniforme acelerado, algunos comentarios pedagógicos}\\
        \footnotesize   \textit{(Velocities: mean, average and instantaneous in uniform accelerated motion, some pedagogical comments)}
}
\author
{ 
Paco Talero $^{1}$, Orlando Organista $^{1}$, Luis H. Barbosa$^{1}$\\
$^{1}$ {\small Grupo F\'{\i}sica y Matem\'{a}tica, Dpt de Ciencias
Naturales, Universidad Central, Bogotá Colombia}
}
\date{}
\twocolumn
[
\begin{@twocolumnfalse}
\maketitle 
\begin{abstract}
Se estudió la relación entre velocidades instantánea, media y promedio en el movimiento unidimensional con aceleración constante. Se demostró que la velocidad instantánea evaluada en el tiempo $t_{p}=\left(t_2+t_1\right)/2$ es igual a las velocidades media y promedio evaluadas entre los tiempos $t_1$ y $t_2$. Se ilustró en detalle la razón por la cual se obtienen  las relaciones antes mencionadas y se usaron los resultados  para  proponer una estrategia pedagógica con el fin de estudiar el movimiento unidimensional con aceleración constante  como una extensión natural del movimiento uniforme rectilíneo.\\
\textbf{Palavras-chave:} velocidad media, velocidad promedio, velocidad instantanea.\\ 

The relation among instantaneous, mean and average velocities in one-dimensional motion with constant acceleration is studied. It was shown that the instant velocity evaluated in the time $t_{p}=\left(t_2+t_1\right)/2$ is similar to the mean and average velocities evaluated between the times $t_1$ and $t_2$.  The reason for relations illustrated before were shown in detail. Also, the results obtained were used to propose a pedagogical strategy in order to study the one-dimensional motion with constant acceleration as a natural extension of one-dimensional motion with constant velocity.
\end{abstract}
\end{@twocolumnfalse}
]

\section{Introducción}

En cursos introductorios de ciencias e ingeniería  es común el estudio del movimiento rectilíneo con aceleración constante (MUA), así como tema obligado en cursos preuniversitarios y de secundaria. Las investigaciones que indagan por el aprendizaje exitoso de  este tema reportan, desde hace casi treinta años y en diversas partes del mundo, que los estudiantes tienen serias dificultades al estudiar el MUA. Tales investigaciones reportan que las principales equivocaciones que cometen  los estudiantes están relacionadas  con: ineficiencia en la operatividad matemática; la inadecuada narración  de  situaciones representadas gráficamente;  la lectura incorrecta  de gráficos simples y la no identificación e interpretación  de áreas y  pendientes \cite{Mc1,Mc2,Beichner}.\\

Un aspecto pedagógico  importante en el estudio  MUA, esencialmente por su sencillez matemática, es el análisis de  las relaciones y diferencias entre las velocidades media,  promedio e instantánea así como su uso en la transposición didáctica  de la cinemática unidimensional \cite{EVOL}. Pero, pese a la gran cantidad de textos que abordan el tema suele no presentarse o presentarse  la ligera\cite{Alonso,Serway,Susan,Hall,ZeM},  desaprovechando de esta manera la riqueza pedagógica  que la relación entre estos conceptos ofrece a la hora de discutir alternativas pedagógicas de aula \cite{EVOL}. \\

En este trabajo se muestra  que para una partícula que tiene un  MUA la velocidad media  es igual a la velocidad promedio y  a su vez estas velocidades son iguales a la velocidad instantánea evaluada en el promedio de los tiempos extremos donde se evaluan tanto la velocidad media como la velocidad promedio. Con base en estas relaciones  se propone una estrategia pedagógica para estudiar el MUA sin el formalismo del cálculo diferencial y extendiendo la idea del movimiento uniforme rectilíneo (MUR).\\

El trabajo está organizado de manera siguiente: en la sección (2) se revisa la definición de las velocidades  media,  promedio e instantánea y se muestran sus relaciones;  en la sección (3) se plantea una propuesta pedagógica que permite abordar  el MUA como una extensión natural del MUR y en la sección (4) se presentan las conclusiones.

\section{Velocidad media,  promedio e instantánea en el MUA}
Desde el punto de vista de la cinemática unidimensional se entiende por partícula  un cuerpo cuya forma y composición interna no afectan de manera significativa la descripción de su movimiento \cite{Landau,EXV}. Así,   se parte  de la hipótesis fundamental  de que se  conoce   -por razones teóricas o experimentales-  ya sea la posición $x(t)$,  la velocidad $v(t)$  o la aceleración $a(t)$ instantáneas de tal partícula como funciones de tiempo.\\  

Para el caso particular de MUA con aceleración $a$ se tienen las expresiones ampliamente conocidas para la posición y la velocidad instantáneas
\begin{equation}\label{pos}
x\left(t\right)=x_{o}+v_{o}t+\frac{1}{2}at^{2},
\end{equation}
\begin{equation}\label{vel}
v\left(t\right)=v_{o}+at.
\end{equation}
La velocidad media $v_{m}$ entre un tiempo $t_1$ y un tiempo $t_2$ se define como 
\begin{equation}\label{velm}
v_m=\frac{x\left(t_2\right)-x\left(t_1\right)}{t_{2}-t_{1}}.
\end{equation}

\begin{figure}[!htp]
\begin{center}
	\begin{tikzpicture}[scale=1.0]
	\path [fill=gray!50] (0cm,0cm) -- (0cm,1cm) -- (4cm,4cm) -- (4cm,0) -- (0,0);
	\draw[-latex, black,line width=1.5pt] (0cm,0cm)--(5cm,0cm);
	\draw[-latex, black,line width=1.5pt] (0cm,0.0cm)--(0cm,5cm);
	\draw[black,line width=1.0pt] (0cm,1cm)--(4cm,4cm);
	\draw [black,dashed,line width=0.8pt] (4cm,4cm) -- (4cm,0cm);
	\draw [black,dashed,line width=0.8pt] (4cm,4cm) -- (0cm,4cm); 
	\coordinate [label=below:\textcolor{black} {$v_{o}$}] (x) at  (-0.25cm,1.2cm);
	\coordinate [label=below:\textcolor{black} {$v$}] (x) at  (-0.25cm,4.25cm);
	\coordinate [label=below:\textcolor{black} {$t$}] (x) at  (4cm,-0.08cm);
	\end{tikzpicture}
\caption{Desplazamiento para el MUA}
\label{gra}
\end{center}
\end{figure}
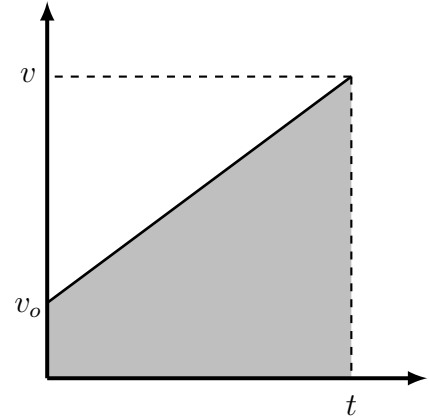

Al reemplazar (\ref{pos}) en (\ref{velm}) se encuentra 
\begin{equation}\label{velmint}
v_{m}=v_{o}+at_{p},
\end{equation}
con
\begin{equation}\label{tmd}
 t_{p}=\frac{t_{2}+t_{1}}{2}.
\end{equation}
Esto significa que la velocidad media entre los tiempos $t_1$ y $t_2$ es igual a la velocidad instantánea evaluada en $t_{p}$, es decir 
\begin{equation}\label{vvm}
v_{m}=v\left(t_{p}\right).
\end{equation}
La velocidad promedio $v_{p}$ entre los tiempos $t_1$ y $t_2$ se define como la semisuma de las velocidades instantáneas $v\left(t_2\right)$ y $v\left(t_1\right)$ \cite{EVOL}. Así
\begin{equation}\label{vvp}
v_{p}=\frac{v\left(t_2\right)+v\left(t_1\right)}{2}.
\end{equation}
En particular cuando $t_1=0$, $t_2=t$, $v(t_1)=v_{o}$ y $v(t_2)=v$  se obtiene 
 
 \begin{equation}\label{vvps}
v_{p}=\frac{v+v_o}{2},
\end{equation}
que  permite calcular el desplazamiento $\Delta x$ como el área del trapecio en la gráfica de velocidad contra tiempo mostrada en la Fig.\ref{gra}, quedando el desplazamiento dado por $\Delta x= v_{p}t$.\\  
Al reemplazar (\ref{vel}) en (\ref{vvp}) se obtiene:
\begin{equation}\label{vpvm}
v_{p}=v_{o}+at_{p}.
\end{equation}
Nótese como  (\ref{vpvm}) implica que la velocidad promedio es igual a la velocidad media y además que estas velocidades son iguales a la velocidad instantánea evaluada en $t_p$.

\section{El MUA: una extensión natural del MUR}

\begin{figure}[!htp]
\begin{center}
	\begin{tikzpicture}[scale=1.0]
	\path [fill=gray!50] (0cm,0cm) -- (0cm,2.5cm) -- (4cm,2.5cm) -- (4cm,0) -- (0,0);
	\draw[-latex, black,line width=1.5pt] (0cm,0cm)--(5cm,0cm);
	\draw[-latex, black,line width=1.5pt] (0cm,0.0cm)--(0cm,5cm);
	\draw[black,line width=1.0pt] (0cm,1cm)--(4cm,4cm);
	\draw[black,line width=1.0pt] (0cm,2.5cm)--(4cm,2.5cm);
	\draw [black,dashed,line width=0.8pt] (4cm,4cm) -- (4cm,0cm);
	\draw [black,dashed,line width=0.8pt] (4cm,4cm) -- (0cm,4cm); 
	\draw [black,dashed,line width=0.8pt] (0cm,1cm) -- (4cm,1cm);
	\draw [black,dashed,line width=0.8pt] (2cm,0cm) -- (2cm,2.5cm);
	\coordinate [label=below:\textcolor{black} {$v_{o}$}] (x) at  (-0.25cm,1.2cm);
	\coordinate [label=below:\textcolor{black} {$v$}] (x) at  (-0.25cm,4.25cm);
	\coordinate [label=below:\textcolor{black} {$t$}] (x) at  (4cm,-0.08cm);
	\coordinate [label=below:\textcolor{black} {$t_p$}] (x) at  (2cm,-0.08cm);
	\coordinate [label=below:\textcolor{black} {$v_{p}$}] (x) at  (-0.25cm,2.8cm);
	\end{tikzpicture}
\caption{Igualdad de desplazamientos en MUA y MUR}
\label{gra2}
\end{center}
\end{figure}
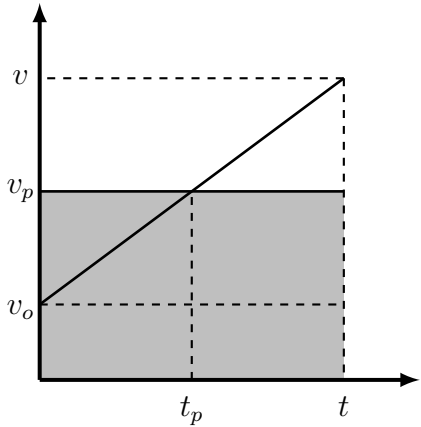

La ventaja pedagógica del uso la velocidad promedio radica en el hecho de que el desplazamiento realizado por una partícula entre los tiempos $t_{1}$ y $t_{2}$ donde tiene velocidades instantáneas $v(t_1)$ y $v(t_2)$ respectivamente, es igual al desplazamiento realizado por una partícula que se mueve con velocidad promedio entre estos mismos tiempos. Esta afirmación se observa al comparar la Fig.\ref{gra} con la Fig.\ref{gra2} donde se ha tomado $t_1=0$, $t_2=t$, $v(t_1)=v_{o}$ y $v(t_2)=v$,  que es  un caso típico abordado en cursos elementales.\\ 

Ahora nótese que, de acuerdo con lo anterior, la ecuación  (\ref{pos}) se puede escribir como
\begin{equation}\label{posp}
x=x_{o}+v_{p}t,
\end{equation}
lo que se interpreta como la posición de la partícula en un tiempo $t$. Resulta entonces clara la analogía con la función de posición contra tiempo de un MUR.\\

La propuesta pedagógica para estudiar el MUA a partir de una extensión natural del MUR, surge del hecho de que es posible iniciar  el estudio del MUR afirmando que se trata del movimiento de una partícula que recorre distancias iguales en tiempos iguales, lo que implica 
\begin{equation}\label{posMUR}
x=x_{o}+v_{o}t,
\end{equation}
 que es análoga a (\ref{posp}). \\
 
Entonces el MUA se puede entender como un movimiento en el cual la partícula en estudio cambia velocidades iguales en tiempos iguales, permitiendo así visualizar el aumento de velocidad o su disminución en cada instante y dando sentido físico a la velocidad promedio. Cabe anotar que referentes concretos que permiten visualizar el cambio de velocidad con el transcurso del tiempo son los ``velocímetros''  de los autos o motocicletas. \\

Por ejemplo: tómese el caso de la caída libre. Desde una cierta altura se deja caer una partícula, si se toma un sistema de referencia positivo hacia abajo se le puede plantear al estudiante como punto de partida de la discusión académica que la velocidad aumenta cada segundo en $10m/s$, en lugar de decirle que la aceleración es constante y de $10m/s^{2}$. Luego se entra a analizar la proporcionalidad que esta afirmación implica junto con las posibles velocidades promedio y sus respectivos desplazamientos. La tabla \ref{tab1} presenta algunos valores de velocidad instantánea, promedio y desplazamiento producida por este método.    

\begin{table}[!htp]
\caption{Ejemplo sobre caída libre}\label{tab1} 
\vspace{2mm}
\centering
\begin{tabular}{|c|c|c|c|}
  \hline
$t\left(s\right)$ & $v\left(m/s\right)$ & $v_p\left(m/s\right)$& $\Delta x\left(m\right)$\\ \hline 
$0,0$  &   $0.0$   & $0,0$   & $0,0$      \\ \hline 
$0,25$  &  $2,5$   & $1,25$   & $0,31$      \\ \hline 
$0,50$  &  $5,0$   & $2,5$   & $1,25$      \\ \hline 
$0,75$  &  $7,5$   & $3,75$   & $2,81$      \\ \hline 
$1,0$  &  $10.0$   & $5,0$   & $5,0$      \\ \hline 
 
\end{tabular}
\end{table}

En general,esto permite entender que una partícula $A$   provista de un MUA que aumentó su velocidad instantánea desde  $v_o$  hasta  $v$ en un tiempo $t$  comparada con  otra partícula $B$ provista de un MUR que mantiene su velocidad  $v_p$  durante todo el tiempo  $t$,  tienen el mismo desplazamiento  y que esto se debe  a que $A$ se mueve más lento que $B$ antes de $t_p$, igual que $B$ justo en $t_p$ y más rápido que $B$ para tiempos mayores que $t_p$.  En otras palabras,el desplazamiento sufrido durante un tiempo $t$ por una partícula provista de un MUA es igual al des\-pla\-za\-mi\-en\-to sufrido por una partícula en MUR con velocidad $v_p$ durante el mismo tiempo.Finalmente, nótese que lo anterior es una aplicación del teorema del valor medio para integrales.  

\section{Conclusiones}
Se presentó una alternativa para abordar el estudio del MUA mediante una extensión del MUR, donde se hace uso exclusivo de cantidades directamente proporcionales. Esto se logró gracias a que se pudo demostrar que la velocidad promedio es igual a la velocidad media y que  además estas velocidades son iguales a la velocidad instantánea evaluada en el tiempo promedio.    

\section*{Agradecimientos}
Los autores agradecen a la Facultada de Ingeniería y al Departamento de Ciencias Naturales de la Universidad  Central por el tiempo y los recursos asignados al proyecto de investigación: Un modelo de enseñanza de la física mediante videos de experimentos Discrepantes realizado durante el año $2013$.   

\renewcommand{\refname}{Referencias}

\end{document}